\documentclass[12pt]{article}

\usepackage{graphicx} 
\begin{document}

\title{Nonlocal generalization of the SM as an explanation of recent CDF result}
\author{N.V.Krasnikov \\ 
Institute for Nuclear Research RAS Moscow, Russia 
\\ 
and
\\ Joint Institute for Nuclear Research, Dubna
}
\maketitle
\begin{abstract}
In this note we show that nonlocal generalization of the standard model (SM)
 with nonlocal scale $\Lambda_{nl} \approx 3~TeV$ 
can explain recent CDF result on the $W$-mass measurement. 
\end{abstract}

\newpage

Recently CDF collaboration has published \cite{CDF} new measured value 
of the $W$-boson mass 
\begin{equation}
m_W = 80.4335 \pm 0.0094~GeV \,
\label{CDF}
\end{equation}
which is in excess of the SM prediction \cite{SM}
\begin{equation}
m_W^{SM} = 80.357 \pm 0.006~GeV \,
\label{SM}
\end{equation}
at $7\sigma$ level. A lot of  explanations of this result has appeared \cite{explanation1} 
- \cite{explanation18}\footnote{Unfortunately the list of references is incomplete}. 

In this note we show that nonlocal generalization of the standard model (SM)
 with nonlocal scale $\Lambda_{nl} \approx 3~TeV$ 
can explain recent CDF result on the $W$-mass measurement. 
Let us illustrate the main idea  of nonlocal field theory 
\cite{Efimov1}-\cite{Efimov5} on the example of the 
$\phi^4$ scalar model with the Lagranjian 
\begin{equation}
L = \frac{1}{2}(\partial^{\mu}\phi\partial_{\mu}\phi - m^2\phi^2) -\lambda\phi^4(x) \,. 
\label{local}
\end{equation}
The Lagrangian (\ref{local}) describes
 renormalizable field theory in which  ultraviolet divergences are elininated by the 
introduction of finite number of local counterterms.
The Lagrangian of the nonlocal analog of  the local  renormalizable $\phi^4$-model (\ref{local}) 
has the form \cite{Efimov1}-\cite{Efimov5}
\begin{equation}
L_{nl} = -\frac{1}{2}\phi(x)(  \partial^{\mu}\partial_{\mu}  +m^2)V^{-1}(-\partial^{\mu}\partial_{\mu})\phi(x)
 -\lambda\phi^4(x) \, 
\label{nonlocal}
\end{equation}
Here the formfactor $ V(p^2)$  is entire function on $p^2$ and 
$ V(p^2) \rightarrow 0$
 at $p^2 \rightarrow -\infty$  that leads to the ultraviolet finite Feynman diagrams. 
Formfactor $ V(z)$ satisfies  the following   requirements \cite{Efimov1}-\cite{Efimov5}:

1. $V(z)$ is entire function on $z$ of the growth $\rho \geq 1/2$.

2. $V(z) \leq  C exp(b|z|^{\rho})$.

3. $V(z) = O(z^{-2})$ at $ Re  ~z \rightarrow -\infty$.

4. $V(z) = O(exp(b|z|^{\rho})$ at $Re ~z \rightarrow \infty$.

5. $V(z) = V(z^*)$.

6. $ V(m^2) = 1$. 
 
 Nonlocal model (\ref{nonlocal}) is unitary and causal. 
For the model (\ref{nonlocal}) the Feynman rules coincide with the 
Feynman rules for local $\phi^4$-model except the propagator 
replacement
\begin{equation}
\frac{1}{p^2 - m^2 +i\epsilon}  \rightarrow \frac{V(p^2)}{p^2 - m^2 +i\epsilon} \,.
\end{equation}
The simplest nonlocal generalization of abelian gauge field(QED)   \cite{Efimov5} consists in the replacement of 
local free photon Lagrangian 
\begin{equation}
L_{B} = - \frac{1}{4}F^{\mu\nu}F_{\mu\nu} \rightarrow L_{nl,B}  = 
-\frac{1}{4}F^{\mu\nu}V^{-1}_{1}(-\partial^{\mu}\partial_{\mu})   F_{\mu\nu} \,,
\label{nlQED}
\end{equation}  
where $ F_{\mu\nu}  = \partial_{\mu}B_{\nu} - \partial_{\nu}B_{\mu}$.
The straightforward  generalization of nonlocal QED to nonabelian gauge theories 
consists in the replacement \cite{krasnikov1987}
\begin{equation}
L_{YM} = -\frac{1}{2}Tr(F^{\mu\nu}F_{\mu\nu}) \rightarrow 
 -\frac{1}{2}Tr(F^{\mu\nu}V^{-1}_{2}(-\Delta^2)    F_{\mu\nu}) \,,
\label{nlQED2}
\end{equation}
where $\Delta^2 = (\partial^{\mu} - igA^{\mu})(\partial_{\mu} - igA_{\mu})   $. 
Nonlocal 
Lagrangian (\ref{nlQED2}) is in fact the generalization of Slavnov 
\cite{Slavnov1, Slavnov2} regularization with 
higher order derivatives. Slavnov regularization corresponds to the 
formfactor $V^{-1}_{Slavnov}(-l^2\Delta^2) = 1 + c_kl^2 (\Delta^2)^k   $. 
A.A.Slavnov has proved \cite{Slavnov1, Slavnov2} that in his regularization all diagrams are 
ultraviolet finite except 
some finite number of diagrams. For instance, for $k = 2$ all diagrams are finite except 
one-loop propagators, three and four vertices.

In this note we  consider   the nonlocal analog of the SM model with nonlocal Lagrangians 
(\ref{nlQED}, \ref{nlQED2})  for $U(1)$ and $SU(2)$ gauge fields $B_{\mu} $ 
and $A_{\mu}$ of the SM model plus 
and it is very essential we use the following nonlocal  generalization of the Higgs field
Lagrangian:
\begin{equation}
L_H = \Delta^{\mu}H^+(V^{-1}_H(-\Delta^{\mu}\Delta_{\mu}))\Delta_{\mu}H \,,
\end{equation}
where $\Delta^{\mu}  = \partial^{\mu} - igA^{\mu} - i\frac{g_1}{2}B_{\mu}$. The existence of the 
nonlocal formfactor $V_H$ leads after electroweak symmetry breaking to 
additional dependence of the $W$- and $Z$-boson masses on $p^2$ momentum, namely 
\begin{equation}
m^2_W \rightarrow m^2_W \times  V^{-1}_H(p^2) \,,
\label{wnonloc}
\end{equation}
\begin{equation}
m^2_Z \rightarrow m^2_Z \times  V^{-1}_H(p^2) \,.
\label{znonloc}
\end{equation}
As a consequence we have the modification of the W-boson and Z-boson propagators, namely 
\begin{equation}
\frac{1}{p^2 - m^2_W } \rightarrow  \frac{1}{p^2 - m^2_W \times  V^{-1}_H(p^2)}  \,.
\end{equation}
\begin{equation}
\frac{1}{p^2 - m^2_Z } \rightarrow  \frac{1}{p^2 - m^2_Z \times  V^{-1}_H(p^2)}  \,.
\end{equation}

Additional formfactor  $V^{-1}_H(p^2)$  leads to the shift of the $W$-boson 
and $Z-boson$ pole masses
\begin{equation}
m^2_W \rightarrow m^2_W \times V^{-1}_H(m^2_W)\,,
\end{equation}
\begin{equation}
m^2_Z \rightarrow m^2_Z \times V^{-1}_H(m^2_Z)\,.
\end{equation}
As a simplest example consider nonlocal formfactor 
\begin{equation}
V_H(p^2) = exp( 4p^2/\Lambda^2_{nl})  \,.
\label{exp}
\end{equation}
Note that the SM prediction for the $W$-boson mass
\begin{equation}
m_W = \frac{g}{G_F^{1/2}}(\frac{1}{(4\sqrt{2})^{1/2}})= m_Z \cos(\theta_W)
\label{Wboson}
\end{equation}
 uses as input parameters  the Fermi constant  $G_F$, the value  $m_Z$ 
of the Z-boson mass and the effective electromagnetic 
coupling constant $\alpha \equiv \alpha(m_Z)$ plus the formulae 
\begin{equation}
g_Z = \frac{g}{cos( \theta_W)} =  2(\sqrt{2}G_F m^2_Z)^{1/2} \,,
\label{Wboson1}
\end{equation} 
\begin{equation}
sin(2\theta_W) = (\frac{4\pi \alpha}{\sqrt{2} G_F m^2_Z})^{1/2}\,.
\label{Wboson2}
\end{equation} 
Using the formulae (\ref{wnonloc}-\ref{Wboson2}) we find that the difference between 
the CDF value (\ref{CDF}) and the SM prediction (\ref{SM}) can be explained 
 for $\Lambda_{nl} \approx 3~TeV$ due to shifts (\ref{wnonloc}, \ref{znonloc}) for W- and Z-boson masses.

In this note for simplicity I  considered  only effect due to nonlocal formfactor  $V_H$ 
for the Higgs field. 
An account of the formfactors $V_1$ and $V_2$ leads to similar effects. The
the sharp change in behaviour of the W- and Z-boson propagators for $p^2 \geq \Lambda^2_{nl}$
takes place. As a consequence we expect the dramatic change 
of the Drell-Yan cross section at the LHC for large invariant $e^+e^-$ and 
$\mu^+\mu^-$  masses.

\end{document}